\documentclass[a4paper]{article}
\pdfoutput=1
\usepackage{authblk}
\usepackage{orcidlink}
\usepackage{graphicx} 
\usepackage{pgffor}
\usepackage{geometry}
\usepackage{booktabs}
\usepackage{amsmath}
\usepackage{xcolor}
\usepackage{hyperref}
\usepackage[sorting=none,style=numeric-comp,giveninits=true,maxnames=1,backend=biber]{biblatex}
\usepackage{xspace} 
\usepackage{upgreek}

\def\MagUp {\mbox{\em Mag\kern -0.05em Up}\xspace}

 \def\Pmu         {\ensuremath{\mu}\xspace}

 \def\Ppi         {\ensuremath{\pi}\xspace}

 \mathchardef\PDelta="7101
 \mathchardef\PXi="7104
 \mathchardef\PLambda="7103
 \mathchardef\PSigma="7106
 \mathchardef\POmega="710A
 \mathchardef\PUpsilon="7107
 \mathchardef\PPi="7105
 \def\PB      {\ensuremath{B}\xspace}                 
 \def\PD      {\ensuremath{D}\xspace}                 

 \def\PK      {\ensuremath{K}\xspace}                 
 \def\Pb      {\ensuremath{b}\xspace}                 
 \def\Pc      {\ensuremath{c}\xspace}

 \def\Ps      {\ensuremath{s}\xspace}

 \def\thebaroffset{0.18em}
\newcommand{\offsetoverline}[2][\thebaroffset]{\kern #1\overline{\kern -#1 #2}}%

\makeatletter
\ifcase \@ptsize \relax
  \newcommand{\miniscule}{\@setfontsize\miniscule{4}{5}}
\or
  \newcommand{\miniscule}{\@setfontsize\miniscule{5}{6}}
\or
  \newcommand{\miniscule}{\@setfontsize\miniscule{5}{6}}
\fi
\makeatother

\DeclareRobustCommand{\optbar}[1]{\shortstack{{\miniscule (\rule[.5ex]{1.25em}{.18mm})}
  \\ [-.7ex] $#1$}}

\def\mup        {{\ensuremath{\Pmu^+}}\xspace}
\def\mun        {{\ensuremath{\Pmu^-}}\xspace} 

\def\mumu       {{\ensuremath{\Pmu^+\Pmu^-}}\xspace}

\def\ellm       {{\ensuremath{\ell^-}}\xspace}
\def\ellp       {{\ensuremath{\ell^+}}\xspace}

\def\squark    {{\ensuremath{\Ps}}\xspace}

\def\cquark    {{\ensuremath{\Pc}}\xspace}
\def\cquarkbar {{\ensuremath{\overline \cquark}}\xspace}
\def\ccbar     {{\ensuremath{\cquark\cquarkbar}}\xspace}
\def\bquark    {{\ensuremath{\Pb}}\xspace}

\def\pion   {{\ensuremath{\Ppi}}\xspace}

\def\pim    {{\ensuremath{\pion^-}}\xspace}

\def\kaon    {{\ensuremath{\PK}}\xspace}

\def\KorKbar {\kern \thebaroffset\optbar{\kern -\thebaroffset \PK}{}\xspace}

\def\Kp      {{\ensuremath{\kaon^+}}\xspace}

\def\Kstarz  {{\ensuremath{\kaon^{*0}}}\xspace}

\def\D       {{\ensuremath{\PD}}\xspace}

\def\DorDbar {\kern \thebaroffset\optbar{\kern -\thebaroffset \PD}\xspace}

\def\Dp      {{\ensuremath{\D^+}}\xspace}
\def\Dm      {{\ensuremath{\D^-}}\xspace}

\def\DpDm    {\ensuremath{\Dp {\kern -0.16em \Dm}}\xspace}

\def\B       {{\ensuremath{\PB}}\xspace}

\def\BorBbar {\kern \thebaroffset\optbar{\kern -\thebaroffset \PB}\xspace}
\def\Bz      {{\ensuremath{\B^0}}\xspace}

\def\Bd      {{\ensuremath{\B^0}}\xspace}

\def\BdorBdbar {\kern \thebaroffset\optbar{\kern -\thebaroffset \Bd}\xspace}

\def\Bs      {{\ensuremath{\B^0_\squark}}\xspace}

\def\BsorBsbar {\kern \thebaroffset\optbar{\kern -\thebaroffset \Bs}\xspace}

\def\Y#1S{\ensuremath{\PUpsilon{(#1S)}}\xspace}

\def\LorLbar     {\kern \thebaroffset\optbar{\kern -\thebaroffset \PLambda}\xspace}

\newcommand{\decay}[2]{\mbox{\ensuremath{#1\!\to #2}}\xspace} 

\def\to                 {\ensuremath{\rightarrow}\xspace}

\def\qsq       {{\ensuremath{q^2}}\xspace}

\def\BdToKpimm    {\decay{\Bd}{\Kp\pim\mup\mun}}

\def\bsll     {\decay{\bquark}{\squark \ell^+ \ell^-}}

\def\AT#1     {\ensuremath{A_{\mathrm{T}}^{#1}}\xspace}           

\def\C#1      {\ensuremath{\mathcal{C}_{#1}}\xspace}                       
\def\Cp#1     {\ensuremath{\mathcal{C}_{#1}^{'}}\xspace}                    
\def\Ceff#1   {\ensuremath{\mathcal{C}_{#1}^{\mathrm{(eff)}}}\xspace}        
\def\Cpeff#1  {\ensuremath{\mathcal{C}_{#1}^{'\mathrm{(eff)}}}\xspace}       
\def\Ope#1    {\ensuremath{\mathcal{O}_{#1}}\xspace}                       
\def\Opep#1   {\ensuremath{\mathcal{O}_{#1}^{'}}\xspace}                    

\newcommand{\aunit}[1]{\ensuremath{\text{\,#1}}}       

\newcommand{\tev}{\aunit{Te\kern -0.1em V}\xspace}
\newcommand{\gev}{\aunit{Ge\kern -0.1em V}\xspace}
\newcommand{\mev}{\aunit{Me\kern -0.1em V}\xspace}
\newcommand{\kev}{\aunit{ke\kern -0.1em V}\xspace}
\newcommand{\ev}{\aunit{e\kern -0.1em V}\xspace}
 
\newcommand{\mevc}{\ensuremath{\aunit{Me\kern -0.1em V\!/}c}\xspace}
\newcommand{\gevc}{\ensuremath{\aunit{Ge\kern -0.1em V\!/}c}\xspace}
\newcommand{\mevcc}{\ensuremath{\aunit{Me\kern -0.1em V\!/}c^2}\xspace}
\newcommand{\gevcc}{\ensuremath{\aunit{Ge\kern -0.1em V\!/}c^2}\xspace}
\newcommand{\gevgevcccc}{\ensuremath{\gev^2\!/c^4}\xspace} 

\def\deriv {\ensuremath{\mathrm{d}}}

\def\gsim{{~\raise.15em\hbox{$>$}\kern-.85em
          \lower.35em\hbox{$\sim$}~}\xspace}
\def\lsim{{~\raise.15em\hbox{$<$}\kern-.85em
          \lower.35em\hbox{$\sim$}~}\xspace}

\def\sPlot{\mbox{\em sPlot}\xspace}

\def\tell1  {TELL1\xspace}
\def\ukl1   {UKL1\xspace}

\newcommand{\lhcborcid}[1]{\href{https://orcid.org/#1}{\hspace*{0.1em}\raisebox{-0.45ex}{\includegraphics[width=1em]{figs/orcidIcon.pdf}}}}

\def\th{{\ensuremath{\theta_h}}\xspace}
\def\tl{{\ensuremath{\theta_\ell}}\xspace}
\def\mkpi{{\ensuremath{m(K\pi)}}\xspace}

\title{Non-parametric and continuous extraction of amplitudes in electroweak penguin decays}
\author{Anja Beck\thanks{anbeck@mit.edu}~~\orcidlink{0000-0003-4872-1213}}
\author{Michele Atzeni\thanks{matzeni@cern.ch}~~\orcidlink{0000-0002-3208-3336}}
\author{Eluned Smith\thanks{eluned@mit.edu}~~\orcidlink{0000-0002-9740-0574}}
\affil{\it\small Department of Physics and Laboratory for Nuclear Science, MIT, Cambridge, 02139, MA, USA}
\date{24th of July 2025}

\begin{document}

\maketitle

\begin{abstract}
We introduce a novel approach to extract the decay-amplitudes in $B\to V(\to M_1M_2)\ellp\ellm$ processes, where $V$ represents a meson with either $J = 0$ (S-wave) or $J = 1$ (P-wave).
This approach enables the decay-amplitudes across the dihadron and dilepton invariant-masses to be extracted from data in a model-independent and continuous way.
To achieve this, the angular decay rate is expressed in a way that allows the application of the \sPlot technique to likelihood fits of the decay angles.
We illustrate the abilities of this method on simulated \BdToKpimm data, containing both S- and P-wave contributions to the $\Kp\pim$ system.  
We provide the weight functions and show that they allow the extraction of the absolute value of the P-wave and S-wave transversity amplitudes as a function of the dihadron and dimuon invariant-masses in a continuous, unbiased, and statistically-powerful way, while only relying on the angular distribution.
As a consequence, the extracted amplitude shapes are model-independent -- up to the angular terms included in the fit -- and can be directly compared to theoretical predictions.
A measurement using this technique can improve the sensitivity to potential new physics effects in \bsll transitions, as well as the understanding of hadronic form factors, particularly in the S-wave system.
\end{abstract}
\vspace{1cm}

Electroweak penguin decays provide an excellent avenue to search for new physics, due to their loop and GIM suppression in the Standard Model (SM). 
In particular, there have been long-standing tensions observed between measurement and prediction in $\Bz\to \Kstarz\mumu$decays~\cite{CMS:2024atz,Belle:2016fev,ATLAS:2018gqc,LHCb-PAPER-2020-002}, which is mediated via the quark-level transition \bsll.
Several other decays and observables of similar \bsll processes have been measured~\cite{LHCb-PAPER-2013-017,LHCb-PAPER-2015-023,LHCb-PAPER-2021-014, LHCb-PAPER-2014-006,LHCb-PAPER-2020-041,LHCb-PAPER-2023-033,LHCb-PAPER-2024-011,CMS:2024syx,CMS:2024atz, BaBar:2012mrf, Belle:2019xld,CMS:2024syx,LHCb-PAPER-2014-024,LHCb-PAPER-2017-013,LHCb-PAPER-2019-040,LHCb-PAPER-2021-004,LHCb-PAPER-2021-038,LHCb-PAPER-2022-045,LHCb-PAPER-2022-046}, many of which exhibit similar tensions to their SM predictions.

The combination of different \bsll measurements leads to sizeable differences in the Wilson coefficient $\mathcal{C}_9$ with respect to its value predicted by the SM~\cite{Altmannshofer:2014rta,Capdevila:2017bsm,Beaujean:2013soa,Descotes-Genon:2013wba,Alguero:2021anc,Alguero:2023jeh}.
Any interpretation of \bsll transitions strongly relies on accurate descriptions of the hadronic physics in each of these decays.
However, modelling these effects and assigning appropriate uncertainties is challenging, complicating a definitive interpretation of the tensions.
In particular, there is suspicion that the deviations may be due to underestimated uncertainties associated with non-local contributions, specifically charm loop processes mediated via $\bquark\to\squark\left[\ccbar\to\gamma^\ast\to\ellp\ellm\right]$,~\cite{Khodjamirian:2010vf,Lyon:2014hpa,Descotes-Genon:2013wba,Ciuchini:2015qxb,Gubernari:2020eft,Gubernari:2022hxn}.
Non-local amplitudes likely have a different dependence on \qsq compared to effects from heavy new physics.
As a consequence, a better understanding of the \bsll \qsq-spectrum is paramount to disentangling these two effects.

Measurements of differential decay rates in \bsll transitions are often performed integrated over ranges in \qsq, see for example Refs.~\cite{LHCb-PAPER-2015-051,LHCb-PAPER-2016-012,LHCb-PAPER-2020-002,LHCb-PAPER-2020-041}.
These are referred to as binned measurements.
While a \qsq-averaged measurement is limited by the finite bin size of typically 1--2\gevgevcccc, it provides information on the angular coefficients -- representing bilinear combinations of the decay amplitudes -- without assumptions on the \qsq shape.

Recently, the LHCb collaboration published measurements of \BdToKpimm decays parametrising the \qsq dependence of the decay amplitudes to allow direct access to the Wilson coefficients~\cite{LHCb-PAPER-2024-011,LHCb-PAPER-2023-033,LHCb-PAPER-2023-032}.
These measurements offer unique data-driven constraints on the size of e.g. non-local contributions, but rely on specific models for the hadronic form factors to describe the decay amplitudes.
As a consequence, the reinterpretation of the measurement using other models may be complicated.

In addition, the composition of the dihadron spectrum  in $B\to V(\to M_1M_2)\ellp\ellm$ decays has strong implications for the complexity of the angular structure and the local $B\to M_1M_2$ form factors.
Often, the dihadron spectrum features a narrow vector-meson resonance (P-wave) on top of a broader structure where $M_1,M_2$ are in a scalar (S-wave) configuration. 
Higher partial waves are also possible.
In the case of \BdToKpimm analyses, a model of the dihadron spectrum is generally employed to better separate the scalar and  $\Kstarz (\to \Kp \pim )$ vector contributions.
The later proposed method is applicable to any set of spin-1 $\Kp\pim$ resonance as they share the same angular structures.
However, we follow existing experimental and theoretical efforts and focus on the $\Kstarz(892)$ state using the shortened notation \Kstarz.
While the $\Kstarz $ dihadron lineshape is typically parametrised using a relativistic Breit-Wigner distribution, the lineshape of the S-wave is largely unknown and can be a significant source of uncertainty on angular coefficients involving the S-wave decay amplitudes.
This paper discusses a new approach for a model-independent continuos extraction of the S-wave dihadron lineshape as well as the variation of the P- and S-wave decay amplitudes across the dimuon invariant-mass squared.

In the following, we first provide a rearranged angular decay rate as well as corresponding weight functions that allow the extraction of the magnitudes squared of the decay amplitudes across the phase space.
Afterwards, we validate our method on both a large toy sample and a toy sample of realistic size.
The paper concludes with a discussion of the advantages and disadvantages of this method with respect to more conventional measurement techniques.

The angular decay rate of \BdToKpimm has been studied in detail, see e.g. Refs.~\cite{Altmannshofer:2008dz, Hofer:2015kka}.
Integrating over the angle between the hadron- and lepton-side decay planes, typically denoted $\phi$, results in the two-dimensional angular decay rate
\begin{align}\label{eq:decrate}
\begin{split}
    \frac{1}{\Gamma_\text{total}}\frac{\deriv\Gamma}{\deriv\cos\th\deriv\cos\tl}
    &= f_1^P(\th,\tl)n_1^P
    + f_0^P(\th,\tl) n_0^P+f_0^S(\th,\tl) n_0^S + f_\beta(\th,\tl)n_\beta \\
    &+\cos ^2\tl \cos \th a_{hc}+\sin^2\tl \cos \th a_{hs}
    +\cos \tl \cos ^2\th a_{\ell c}+\cos \tl \sin^2\th a_{\ell s} \ , \\
\end{split}
\end{align}
with the angular functions
\begin{align}\label{eq:angularfunc}
\begin{split}
    f_1^P(\th,\tl) &= \frac{9}{32} \left(\cos ^2\tl+1\right) \sin^2\th \ , \\
    f_0^P(\th,\tl) &= \frac{9}{8} \sin^2\tl \cos ^2\th \ , \\
    f_0^S(\th,\tl) &= \frac{3}{8} \sin^2\tl \ , \\
    f_\beta(\th,\tl) &= \frac{3}{8}r_\beta\sin^2\th+\frac{3}{4}(1-r_\beta)\cos^2\th  \ .
\end{split}
\end{align}
The coefficients represent combinations of the transversity amplitudes,
\begin{align}\label{eq:coeffs}
\begin{split}
n_1^P &= \beta ^2 \left(| A_{\parallel}^L| {}^2+| A_{\perp}^L| {}^2+| A_{\parallel}^R| {}^2+| A_{\perp}^R| {}^2\right) \\
n_0^P &= \beta ^2 \left(| A_0^L| {}^2+| A_0^R| {}^2\right) \\
n_0^S &= \beta ^2 \left(| {A'}_0^L| {}^2+| {A'}_0^R| {}^2\right) \\
n_\beta r_\beta &= \frac{(1-\beta ^2)}{8} \left(
3 \left(\beta ^2+2\right) \Re\left[A_{\parallel}^L (A_{\parallel}^R){}^*\right]
+3\beta ^2 \Re\left[A_{\perp}^L (A_{\perp}^R){}^*\right]
+6 \Re\left[A_{\perp}^L (A_{\perp}^R){}^*\right]\right. \\
&\left.\qquad\qquad\qquad+8 \Re\left[{A'}_0^L ({A'}_0^R){}^*\right]+4 | {A'}_t| {}^2\right)
+\frac{(1-\beta ^2)}{4 \beta ^2} (3n_1^P+2 n_0^S) \\
n_\beta (1-r_\beta) &= \frac{(1-\beta ^2)}{4} \left(
2\Re\left[{A'}_0^L ({A'}_0^R){}^*\right]
+6\Re\left[A_0^L(A_0^R){}^*\right]\right. \\
&\left.\qquad\qquad\qquad
+| {A'}_t| {}^2
+3| A_S| {}^2
+3| A_t| {}^2\right)
+\frac{(1-\beta ^2)}{4\beta^2} \left(3 n_0^P+n_0^S\right) \\
\end{split}
\end{align}
and
\begin{align}\label{eq:asym}
\begin{split}
a_{hc} &= \frac{3}{8} \sqrt{3} \left(1-\beta ^2\right)
\Re\left[\left(A_0^R\right){}^* {A'}_0^L+\left(A_0^L\right){}^* {A'}_0^R+\left(A_t\right){}^* {A'}_t+\left(A_0^L\right){}^* {A'}_0^L+\left(A_0^R\right){}^* {A'}_0^R
\right] \\
a_{hs} &= \frac{3}{8} \sqrt{3} \left(\left(1-\beta ^2\right)\Re\left[\left(A_0^R\right){}^* {A'}_0^L+\left(A_0^L\right){}^* {A'}_0^R+\left(A_t\right){}^* {A'}_t\right]\right. \\
&\left.\qquad\quad+\left(1+\beta ^2\right) \Re\left[\left(A_0^L\right){}^* {A'}_0^L+\left(A_0^R\right){}^* {A'}_0^R\right]\right) \\
a_{\ell c} &= \frac{9}{16} \beta \left(1-\beta ^2\right) \Re\left[\left(A_S\right){}^* \left(A_0^L+A_0^R\right)\right] \\
a_{\ell s} &= \frac{9}{8} \beta \Re\left[A_{\parallel}^L \left(A_{\perp}^L\right){}^*-A_{\parallel}^R \left(A_{\perp}^R\right){}^*\right]
\end{split}
\end{align}
where $\beta=\sqrt{1-\tfrac{4m_\ell^2}{\qsq}}$ is the speed of the muons in the dimuon restframe and $A$ and $A'$ are amplitudes belonging to the P-wave and S-wave contributions respectively.
The subscript denotes the transverse, $\parallel,\perp$, longitudinal, $0$, time-like, $t$, or scalar, $S$, nature of the amplitude and the superscript indicates whether the amplitude stems from a left- ($L$) or right-handed ($R$) current.
The amplitudes carry implicit dependence on the dihadron invariant mass \mkpi and the dilepton invariant-mass squared \qsq. 
Note that the lineshape in the dihadron invariant-mass of the S- and P-wave have been absorbed into the amplitudes for better readability.
The integral over the angular decay rate imposes the normalisation condition for the coefficients
\begin{align}
    \frac{1}{\Gamma_\text{total}}\int\int\frac{\deriv\Gamma}{\deriv\cos\th\deriv\cos\tl}~\deriv\cos\th~\deriv\cos\tl = n^P_0+n^P_1+n^S_0+n_\beta = 1 \ .
    \label{eq:norm}
\end{align}

The decay rate in Eq.~\eqref{eq:decrate} is expressed using an unconventional angular basis where the four coefficients $n^P_{0,1}$, $n^S_0$, and $n_\beta$ are strictly positive and associated with linearly independent angular functions.
This allows for the construction of a set of weight functions
\begin{align}\label{eq:weights}
\begin{split}
w_1^P(\th,\tl) &= -\frac{1}{2}\left(\left(5 \cos^2\th-3\right)+\frac{r_\beta}{1-r_\beta}\left(5 \cos^2\th-1\right)\right)\left(5 \cos^2\tl-1\right) \ , \\
w_0^P(\th,\tl) &= \frac{1}{4}\left(\cos^2\th \left(35-100 \cos^2\tl\right)+30 \cos^2\tl-11\right)+\frac{r_\beta}{8 (1-r_\beta)}\left(5 \cos^2\th-1\right) \left(5 \cos^2\tl-1\right) \ , \\
w_0^S(\th,\tl) &= \frac{3}{4} \left(5 \cos^2\th-3\right) \left(5 \cos^2\tl-2\right)-\frac{3 r_\beta }{8 (1-r_\beta)}\left(5\cos^2\th-1\right) \left(5 \cos^2\tl-1\right) \ , \\
w_\beta(\th,\tl) &= \frac{3}{4 (1-r_\beta)} \left(5 \cos^2\th-1\right) \left(5 \cos^2\tl-1\right) \ .
\end{split}
\end{align}
Each weight function enables the extraction of the shape of the corresponding coefficient from data.
This method is known as the \sPlot technique~\cite{Pivk:2004ty}.
It is a commonly used tool in high-energy physics and is almost exclusively employed to separate signal from background using the final-state invariant-mass as a discriminator.
The non-traditional angular basis in Eq.~\eqref{eq:decrate} allows for a novel application of the \sPlot technique using the decay angles as the discriminant variables in order to extract the shape of the magnitude of the transverse and longitudinal P-wave amplitudes, via $n_1^P$ and $n_0^P$, the S-wave amplitude, via $n_0^S$, and a more complex combination of amplitudes suppressed by $(1-\beta)$, via $n_\beta$, in a continuous, data-driven way.

In order to illustrate and test our method, toy data is generated using the parametrisation described in Refs.~\cite{LHCb-PAPER-2023-033,LHCb-PAPER-2023-032}.
The Wilson coefficients are set to a SM prediction obtained using the \texttt{eos} software~\cite{EOSAuthors:2021xpv} and the non-local hadronic contributions are neglected.
The values for the local P-wave form factor coefficients are taken directly from Ref.~\cite{Gubernari:2022hxn} and the S-wave lineshape is described using a LASS model~\cite{Aston:1987ir,LHCb-PAPER-2023-033,LHCb-PAPER-2023-032}.

The angular distribution in Eq.~\eqref{eq:decrate} captures all structures that can be generated by spin-0 and spin-1 resonances decaying to $\Kp\pim$, where the P-wave is interpreted as the \Kstarz resonance.
Contributions from other spin-1 states are possible and can complicate the interpretation of the observables.
Experimental results indicate the presence of D-wave states~\cite{Belle:2014nuw,LHCb-PAPER-2016-025} generating higher order angular terms which may bias the fit result.
Due to the strong suppression of the D-wave below $\mkpi \sim 1\gevcc$~\cite{Belle:2014nuw} however, its contribution can often be neglected.
Despite considering only S- and P-wave contributions, this work studies the \mkpi range up-to 1.5\gevcc in order to show the variation of the full S-wave lineshape.

\begin{figure}[t!]
    \centering
    \includegraphics[width=.5\textwidth]{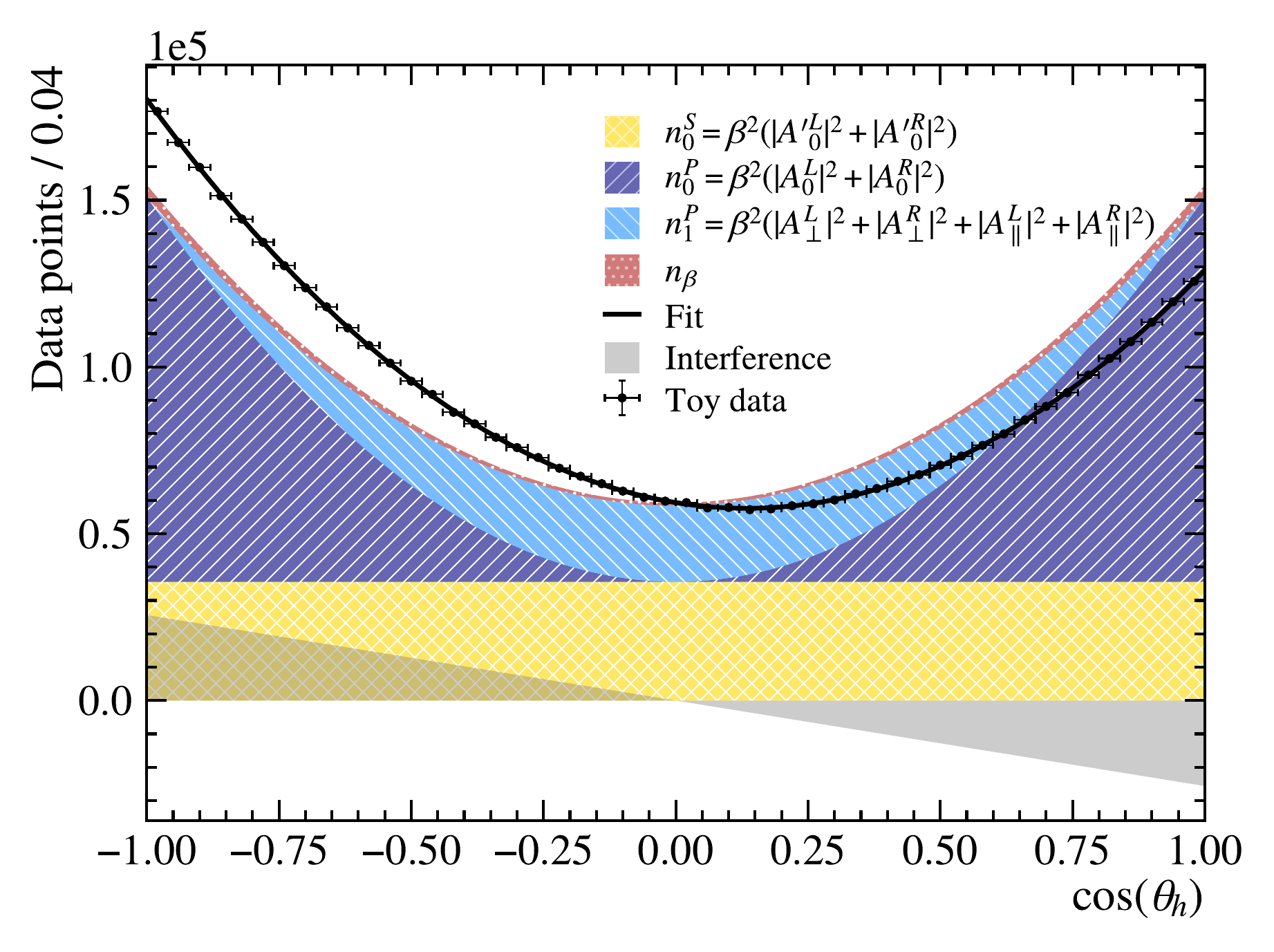}%
    \includegraphics[width=.5\textwidth]{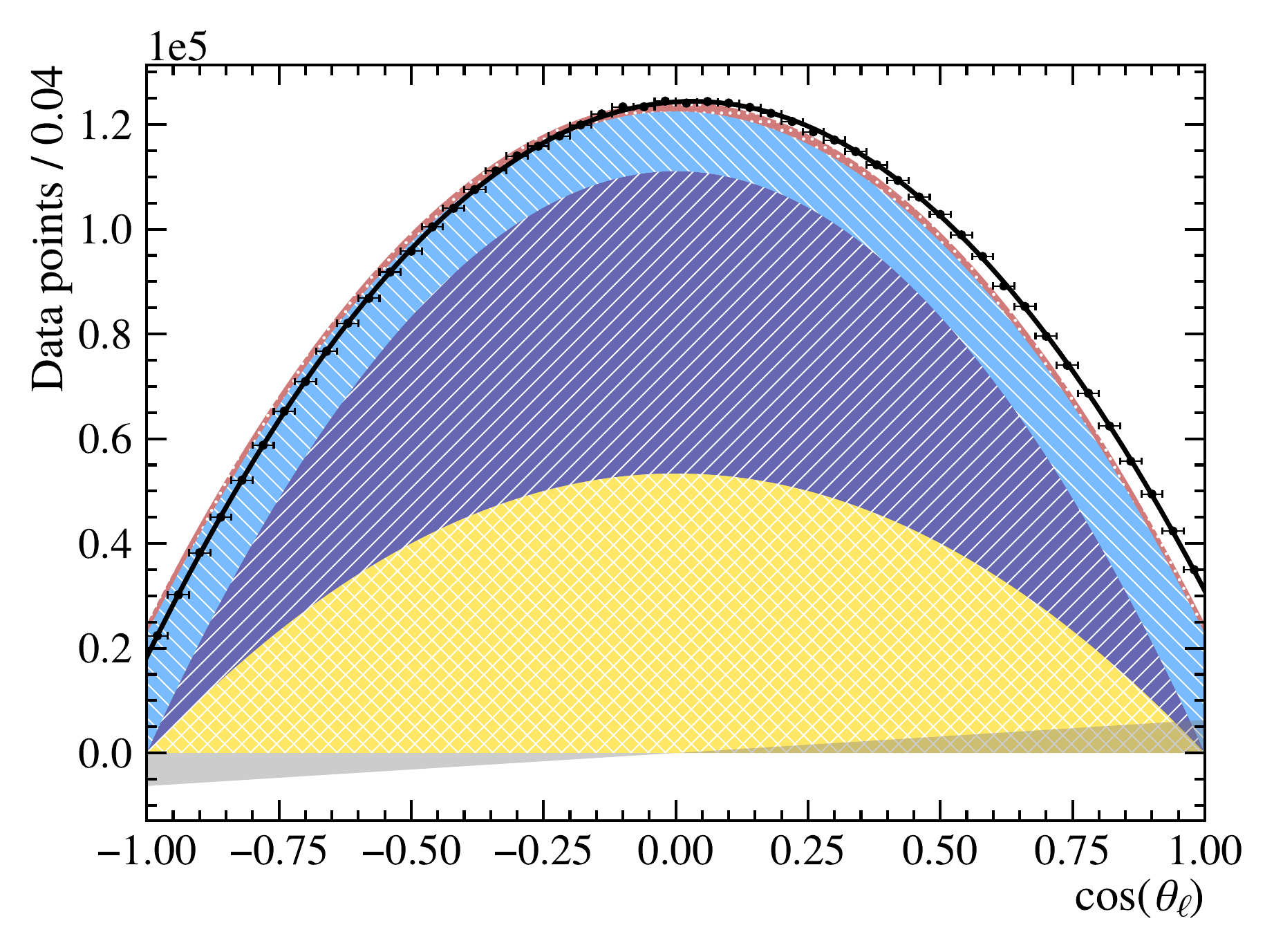}
    \caption{Projection of a large toy data set onto the (left) hadron helicity angle and (right) muon helicity angle. The fitted angular decay rate including all its components is also shown.}
    \label{fig:fitprojections}
\end{figure}

Figure~\ref{fig:fitprojections} shows the fit to a large toy data set with 4.5 million generated decays within the ranges $\mkpi<1.5\gevcc$ and $1.1<\qsq<7\gevgevcccc$, projected onto the two angles.
For numerical stability, the two products $n_\beta r_\beta$ and $n_\beta(1-r_\beta)$ are used as fit parameters instead of the variables $n_\beta$ and $r_\beta$.
Note that the interference in the hadron helicity angle (left plot) originates from interference between S- and P-wave amplitudes while the interference in the muon helicity angle (right plot) originates from interference between P-wave amplitudes.

\begin{figure}[b!]
    \centering
    \includegraphics[width=.5\textwidth]{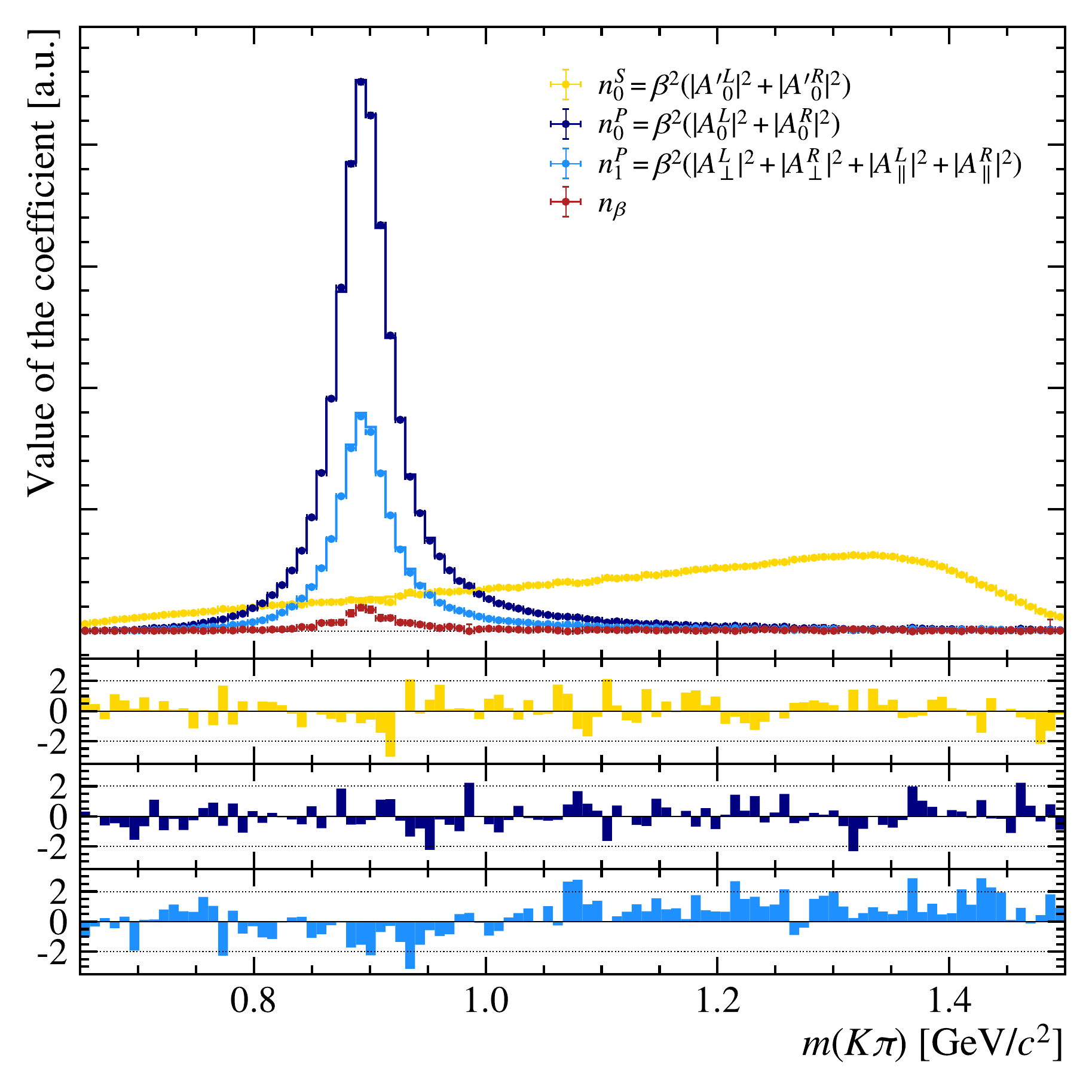}%
    \includegraphics[width=.5\textwidth]{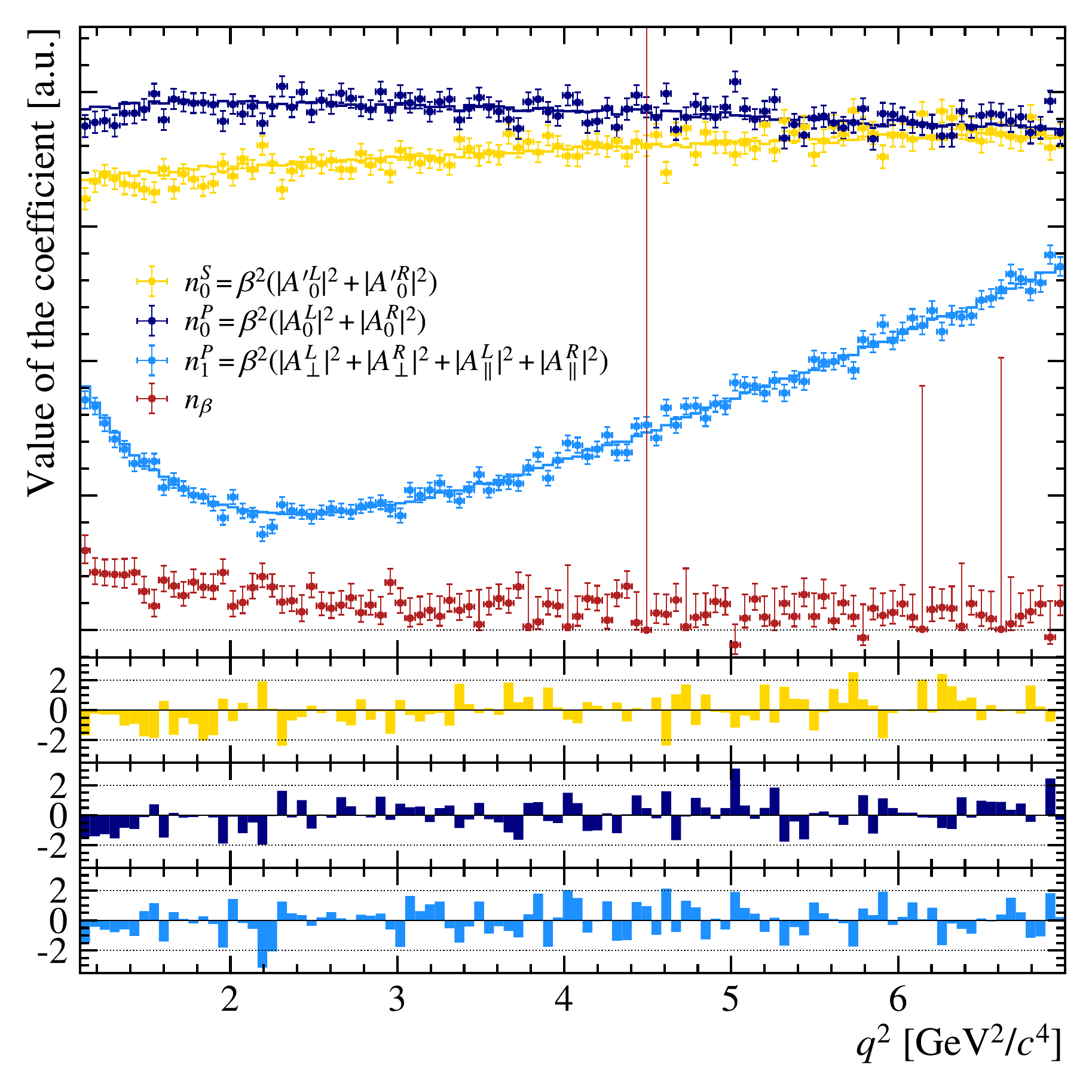}
    \caption{Extracted shapes (data points) compared to the generated shapes (lines) in the (left) dihadron invariant-mass and (right) dimuon invariant-mass squared for a large toy data set.
    The three bottom plots show the bin-wise pulls for the three components in their respective colour.
    No generated shape for the $(1-\beta)$-dependent contribution is shown because the generation of a representative toy sample is non-trivial.}
    \label{fig:proof}
\end{figure}
Using the fitted angular distribution, the \sPlot technique~\cite{Pivk:2004ty} allows to calculate weights that extract the individual components.
Figure~\ref{fig:proof} compares the extracted (data points) shapes from the same large toy with 4.5 million decays to generated (lines) distributions obtained by sampling from a model where only the amplitudes of interest are non-zero.
By construction, the dependence on the dihadron invariant-mass is identical for the two P-wave coefficients $n_i^P$ and corresponds to the Breit-Wigner model while the dependence for the S-wave contribution $n_0^S$ follows the LASS shape and the $(1-\beta)$-suppressed coefficient is a combination of both.
Because the $(1-\beta)$-dependent contribution contains products of different amplitudes, it is difficult to produce the distribution using our toy generator and no expectation is provided.
The pulls in the bottom part of the figures show that this novel application of the \sPlot technique provides an unbiased proxy for the absolute values of the transversity amplitudes across the phase space including correct relative scaling between them.
Despite the large size of the data set (4.5 million decays to determine five parameters in a two-dimensional fit), a small residual fit instability related to the $n_\beta$ contribution remains.
Due to this instability and the large correlation between the $n_\beta$ and $n_1^P$ terms, the latter exhibits a small bias above $1\gevcc<\mkpi$.
A fit to a data set of realistic size, see below, requires additional constraints to ensure stability while also resulting in larger statistical uncertainties.
As discussed below, this leads to no observable bias in the coverage.
Moreover, the aim of this method is to extract the P-wave and the S-wave shapes which is anyhow unbiased via $n_0^P$ and $n_0^S$.
Note that the interference terms shown in gray in Fig.~\ref{fig:fitprojections} vanish in the projections on the invariant-masses due to the asymmetry in the angular terms.

For the sake of illustrating the expected statistical uncertainty, and to demonstrate the feasibility of the technique on a more realistic sample-size, we assume 45~000 \BdToKpimm candidates in the ranges $\mkpi<1.5\gevcc$ and $1.1<\qsq<7\gevgevcccc$.
This is a rough estimate of the size of the data set that could be collected by the LHCb collaboration by the end of the LHC Run 3 data-taking period, extrapolated from previous measurements, e.g. Ref.~\cite{LHCb-PAPER-2020-002}.
In order to achieve a stable fit, either the fit coefficient $n_\beta r_\beta$ or $n_\beta (1-r_\beta)$ must be fixed or constrained to a known value such as the SM prediction.
Due to the small magnitude of $n_\beta$, the systematic uncertainty associated with this choice is negligible in the considered scenario.

\begin{figure}
    \centering
    \includegraphics[width=.5\textwidth]{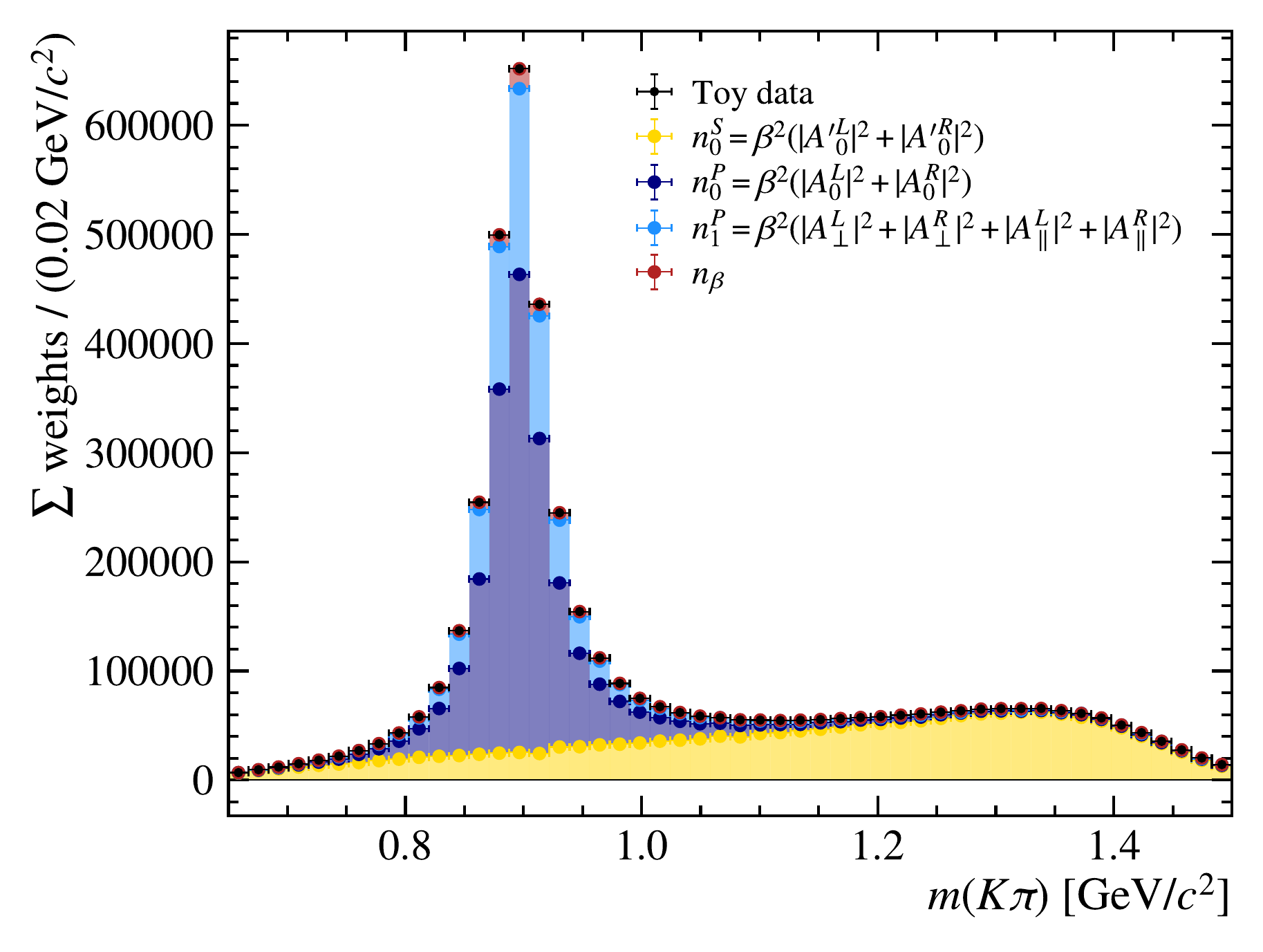}%
    \includegraphics[width=.5\textwidth]{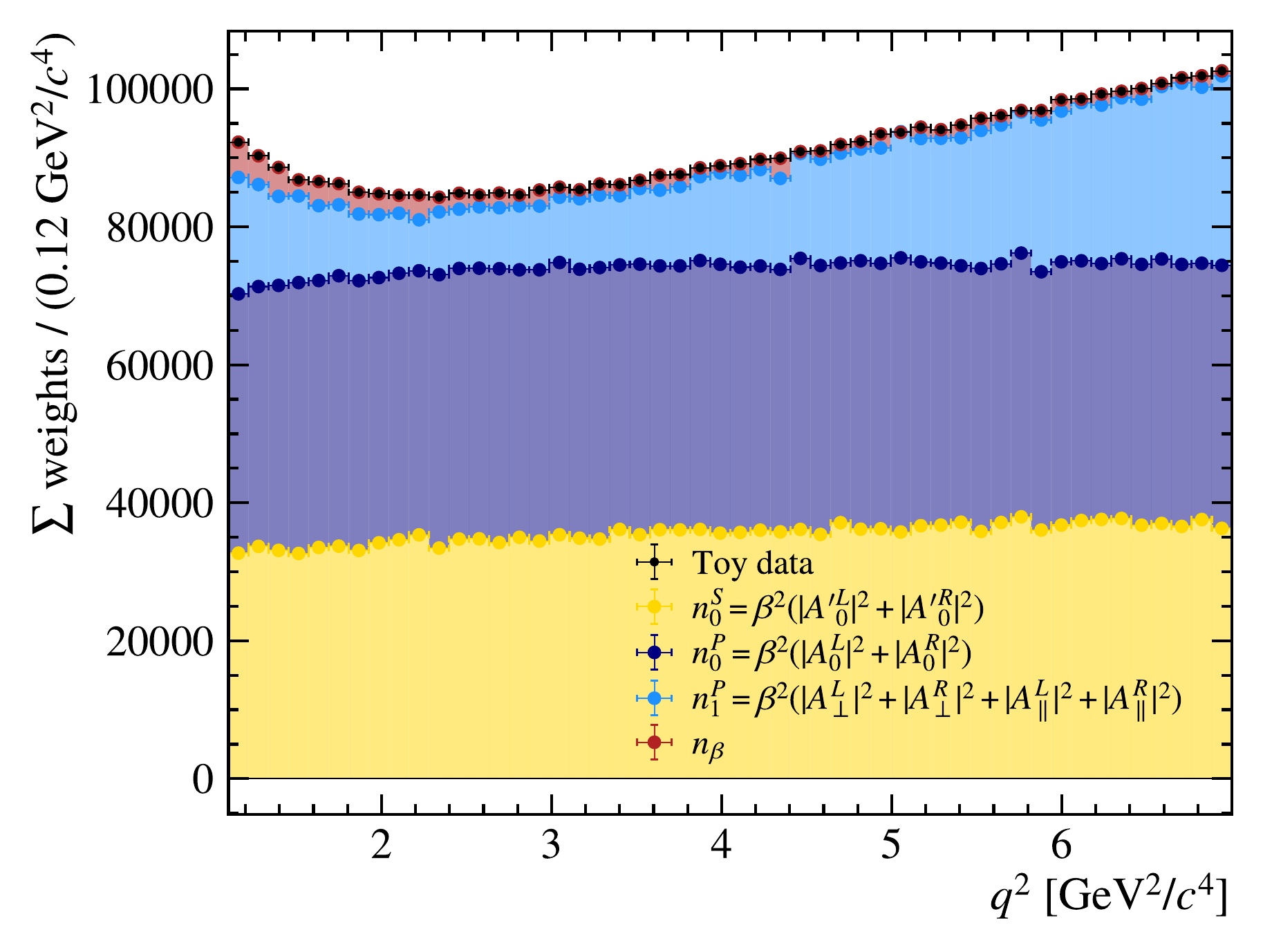}
    \caption{Expected statistical uncertainty on the shape of the four components in the (left) dihadron and (right) dimuon invariant-mass based on the chosen binning scheme and a data set of 45~000 \BdToKpimm decays.}
    \label{fig:sweighted_unc}
\end{figure}

Figure~\ref{fig:sweighted_unc} shows the resulting dimuon invariant-mass squared and dihadron invariant-mass distributions for the four components $n_i$ including their estimated statistical uncertainty based on the assumed number of candidates.
The uncertainty is calculated from the sum of weights squared in every bin.
The coverage of these uncertainties is tested using toy studies.
The angular fit is performed for each toy and the different components are extracted based on the resulting weights.
In a two-dimensional histogram using \mkpi and \qsq, the value and uncertainty -- represented by the sum of weights and the square root of the sum of weights squared -- in every bin is determined.
Coverage is confirmed by examining the pull distribution as well as its mean and standard deviation in every bin.
No bias or faulty coverage is observed locally or globally.
In the chosen scenario, the statistical uncertainties exceed the magnitude of the $(1-\beta)$-suppressed contribution in all bins.
As a consequence, fixing or constraining the $(1-\beta)$-suppressed contributions to a slightly wrong value -- or even setting them to zero -- results in a small local bias.
This bias is likely shared between the three components and hence limited to a fraction of the statistical uncertainty.

\begin{figure}
    \centering
    \includegraphics[width=.5\textwidth]{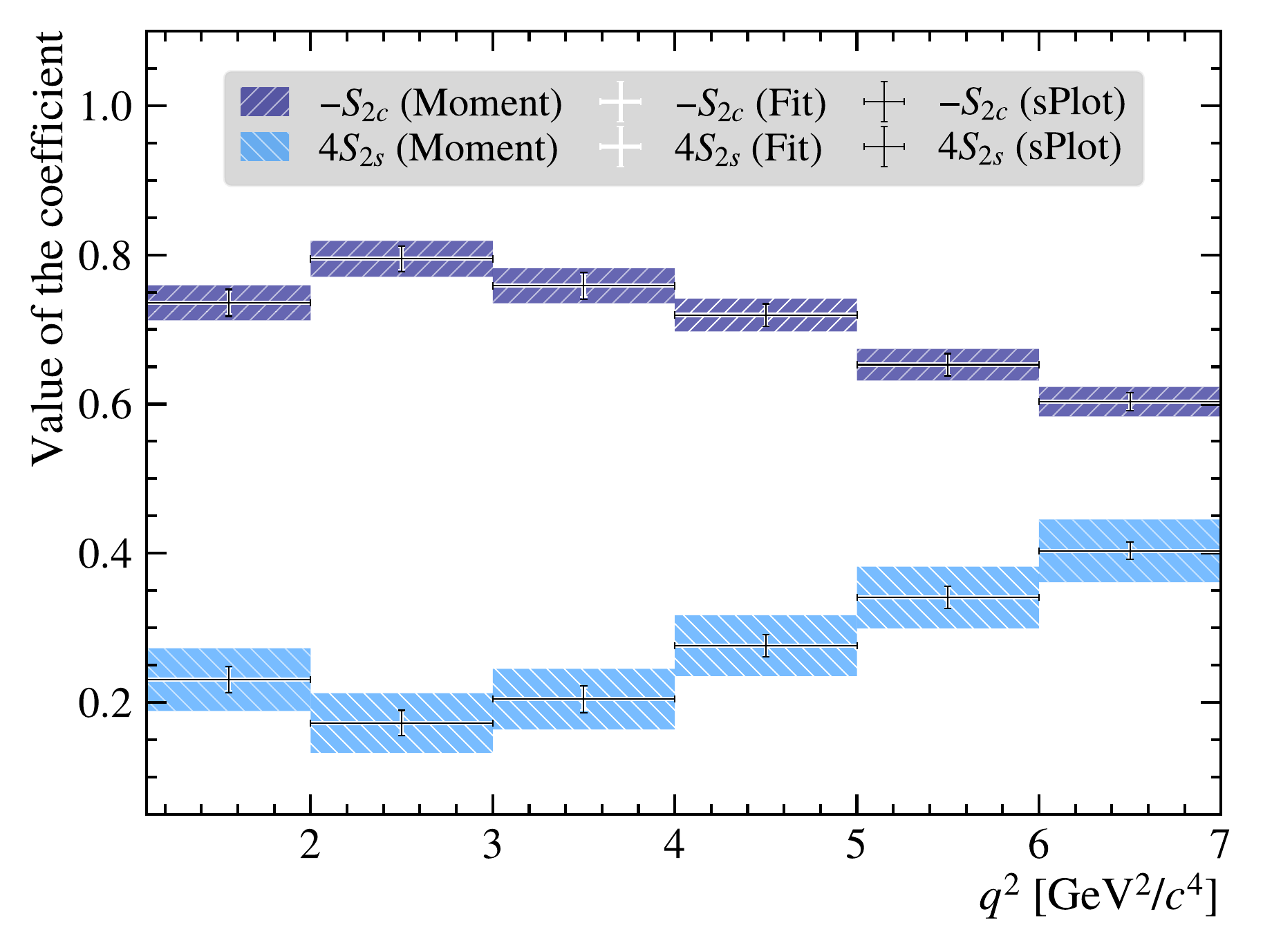}%
    \includegraphics[width=.5\textwidth]{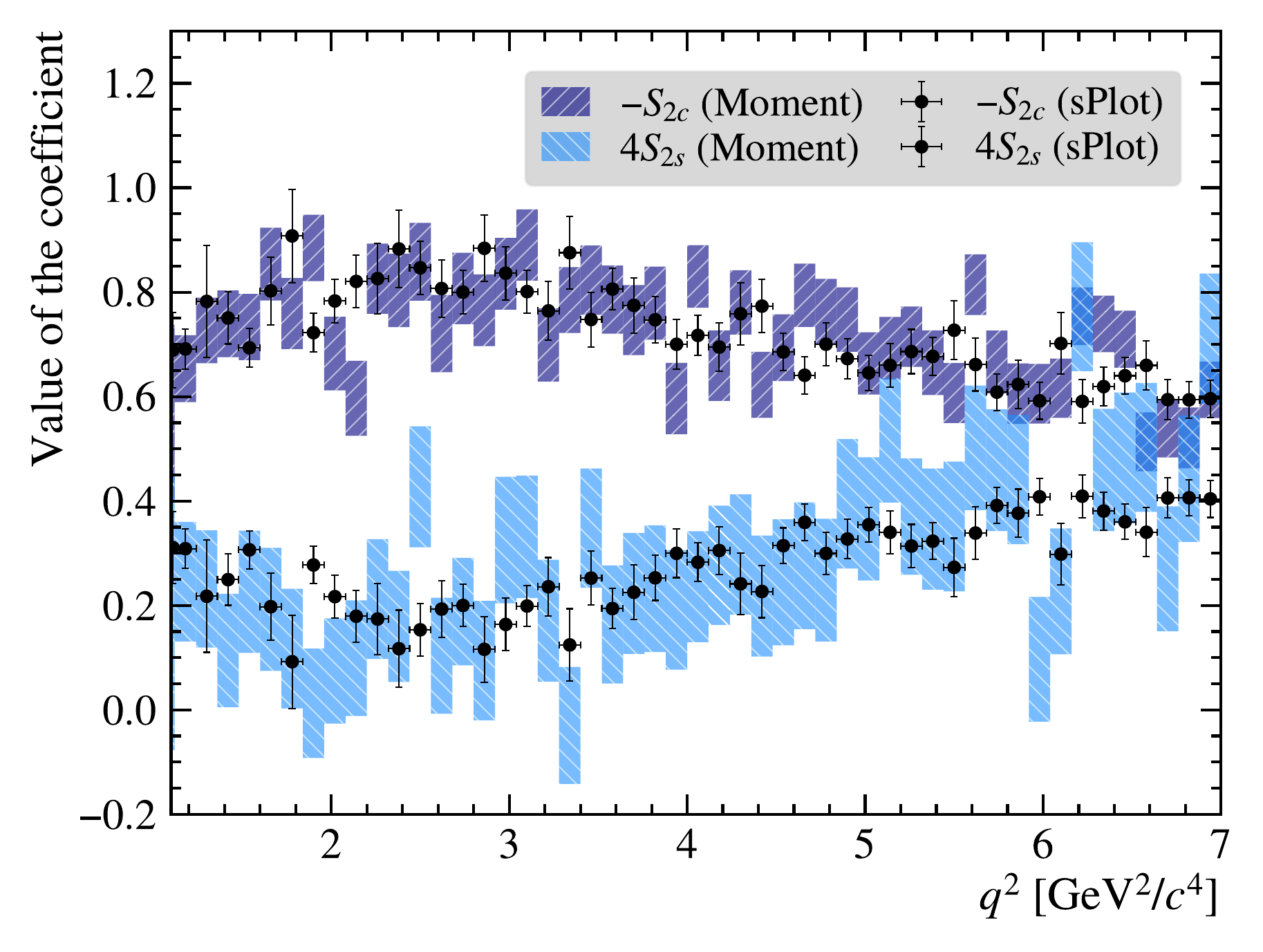}
    \caption{Values of the two angular coefficients $-S_{2s}$ and $4S_{2c}$ across the dimuon invariant-mass squared including the statistical uncertainty when extracted using the method of moments (coloured bands), an angular fit per bin (white markers), and the \sPlot technique (black markers).
    All uncertainties are estimated using data sets containing 45~000 decays representing P-wave only for the moments and a mix of P- and S-wave for the fit-based methods.
    In the left plot the central value is fixed to the generated value for easier comparison of the statistical uncertainties, and a standard bin width of 1 \gevgevcccc is used.
    The much finer binning scheme in the right plot showcases the reduced scatter of the \sPlot technique compared to the method of moments.
    No results of per-bin likelihood fits are shown in the finer binning scheme as the limited amount of data in each bin prevents stable per-bin likelihood fits.}
    \label{fig:sweighted_unc_moments}
\end{figure}

The coefficients appearing in Eq.~\eqref{eq:decrate} can be mapped to the more conventional $S_i$ angular observables~\cite{Altmannshofer:2008dz}, for example
\begin{align}
    \left(\frac{\deriv\Gamma}{\deriv\qsq}\right)^{-1}n_0^P = -S_{2c} \quad\text{and}\quad \left(\frac{\deriv\Gamma}{\deriv\qsq}\right)^{-1}n_1^P = 4 S_{2s} \ .
\end{align}
Figure~\ref{fig:sweighted_unc_moments} shows these two angular coefficients.
The difference in shape compared to the previous figures stems from the additional normalisation to the decay rate as a function of \qsq.
For comparison, the figure includes the statistical uncertainty on the same coefficients extracted using (i) the method of moments~\cite{Beaujean:2015xea} (ii) from the result of maximum-likelihood fits of Eq.~\eqref{eq:decrate}.
The \sPlot technique and the per-bin likelihood fits are performed on a mixture of S- and P-wave on a data sample with 45~000 candidates.
The uncertainties obtained from these fit-based techniques have similar magnitude and are consistent between the two $S_i$ observables.\footnote{
Of course, fitting Eq.~\eqref{eq:decrate} to the full sample results in smaller uncertainties \textit{on the fit parameters} compared to the fits in bins.
This uncertainty is however not directly related to the calculation of the \sPlot weights given in Eq.~\eqref{eq:weights}.
The per-bin uncertainty associated with the \sPlot method shown in Fig.~\ref{fig:sweighted_unc_moments} is dominated by the amount of data per bin resulting in similar precision compared to the fit in each bin.}
In contrast, the angular moments are calculated on a data set of the same size but without the S-wave contribution because there is no trivial way to separate the two components in a moment analysis.
As a consequence, any conclusion about the comparative statistical power between the moments and the other two methods is very conservative.
In order to allow a direct and easy comparison of the different uncertainties, the central values in the left part of Fig.~\ref{fig:sweighted_unc_moments} are fixed to the generated ones.
In contrast, the right part of Fig.~\ref{fig:sweighted_unc_moments} uses a finer binning scheme to showcase the reduced fluctuations in the central values as a consequence of the inherent correlation across the phase space exploited by the technique proposed in this paper.
The results of the per-bin likelihood fits are omitted in this case given the poorer fit stability due to the limited amount of data in each bin.

Given the nature of the studies presented here, a result cannot easily be listed as numbers or an analytic expression.
The results of a measurement using this technique may be represented as histograms.
This scenario is likely the easiest to implement as the uncertainties in each bin as well as the inter-bin correlations are clearly defined.
However, an alternative way of representing the extracted distributions in an unbinned fashion is via individual unweighted data sets generated using dynamic nested sampling~\cite{Higson_2018} as implemented for example in the \textit{dynesty} package~\cite{sergey_koposov_2024_12537467}.
Another option is to express the distributions with normalising flows using the \textit{nabu-hep} tool~\cite{Araz:2025ezp} designed for this purpose specifically.
Finally, the result may of course also be presented by publishing the data including the extracted weights (or an ensemble of extracted weights obtained from bootstrapping the original data) if possible and useful.
The presentation of uncertainties in a continuous way and the correct accounting for correlations across the phase space is particularly challenging.
The development and study of different ideas and methods is outside the scope of this paper and may be examined in future work.

As evidenced by Fig.~\ref{fig:proof}, the results of the \sPlot technique when applied to the angular decay rate results in an unbiased extraction of the angular coefficients $n_i$.
Moreover as illustrated by Fig.~\ref{fig:sweighted_unc_moments}, the resulting statistical uncertainty is smaller compared to the method of moments and similar to per-bin likelihood fits.
As a direct consequence of the correlations across the phase space, the scatter around the central value is reduced.
Another advantage over the method of moments is the ability to extract the shape of the S-wave without assuming massless leptons.
Additionally, the \sPlot method proposed here has a major advantage over per-bin likelihood fits as the width of the bins in the latter is limited by the amount of data required to achieve a stable fit.
Results obtained using the presented technique would enable thorough studies of the \qsq-dependence of the different amplitudes.
What is more, better knowledge of the S-wave dihadron lineshape can reduce systematic uncertainties present in experimental measurements and help improve the understanding on $B\to M_1M_2$ form factors across the full dihadron invariant-mass range.

A drawback of our method is that the angular coefficients can be strongly correlated which can lead to difficulties in the likelihood fit.
This is particularly problematic for the $(1-\beta)$ suppressed terms.
Transformations of the $(1-\beta)$ dependent terms in the decay rate and the inclusion of a wider \qsq range, as well as the addition of background and other realism, may be the focus of future investigations.
Finally as $A_\perp$ and $A_\parallel$ only appear together in Eq.~\eqref{eq:decrate}, this method can only provide a combined shape for these two amplitudes.

In summary, a model-independent method to extract the dimuon and dihadron invariant-mass shapes of amplitudes appearing in \BdToKpimm decays has been presented.
The dependence on the dimuon invariant-mass squared is particularly interesting for the study of the underlying \bsll structure and potential deviations from the Standard Model expectation.
The shapes of the contributions to the dihadron spectrum can be used to further reduce the systematic uncertainty on experimental measurements and provide better understanding of the QCD resonance spectrum.
Note that this method can be applied to any $B\to V(\to M_1M_2)\ellp\ellm$ decay and translated to other spin configurations after establishing an appropriate expression for the angular decay rate.

\section*{Acknowledgements}
We would like to thank the US National Science Foundation, whose funding under award number 2310073 has helped support this work. 

\printbibliography

\end{document}